\RequirePackage[l2tabu, orthodox]{nag}
\RequirePackage{snapshot}

\documentclass[11pt,onecolumn]{article}

\makeatletter
\if@twocolumn
  \usepackage[dvips,letterpaper,top=0.5in, bottom=0.5in, left=0.75in, right=0.5in,includefoot,heightrounded]{geometry}
\else
  \usepackage[dvips,letterpaper,margin=1in,includefoot,heightrounded]{geometry}
\fi

\usepackage{srcltx}

\usepackage{amsmath}
\usepackage{amssymb,amsfonts}

\usepackage{abstract}

\usepackage{graphicx}
\usepackage{epsfig}
\usepackage[usenames,dvipsnames]{color}
\usepackage[usenames,dvipsnames]{xcolor}
\usepackage{subfigure}

\usepackage{booktabs}

\usepackage{setspace}
\usepackage{flushend}
\usepackage{multicol}

\usepackage{cite}
\usepackage{url}
\usepackage[normalem]{ulem}

\usepackage{enumerate}

\usepackage{multirow}

\usepackage{hyperref}

\usepackage{lipsum}

\usepackage[sc,tiny,center]{titlesec}

\usepackage{microtype}

\makeatletter

\newenvironment{rsmallmatrix}{\null\,\vcenter\bgroup
  \Let@\restore@math@cr\default@tag
  \baselineskip6\ex@ \lineskip1.5\ex@ \lineskiplimit\lineskip
  \ialign\bgroup\hfil$\m@th\scriptstyle##$&&\thickspace\hfil
  $\m@th\scriptstyle##$\crcr
}{%
  \crcr\egroup\egroup\,%
}
\makeatother

\title{%
An Orthogonal 16-point Approximate DCT for Image and Video Compression
}

\author{T. L. T. da Silveira%
\thanks{%
T. L. T. da Silveira
is with the
Programa de P\'os-Gradua\c{c}\~ao em Inform\'atica,
Universidade Federal de Santa Maria,
Santa Maria, RS, Brazil,
\protect\url{thiago@inf.ufsm.br}
}
\and
F.~M.~Bayer%
\thanks{%
F. M. Bayer is with the
Departamento de Estat\'istica
and LACESM,
Universidade Federal de Santa Maria,
Santa Maria, RS, Brazil,
\protect\url{bayer@ufsm.br}}
\and
R.~J.~Cintra%
\thanks{%
R.~J.~Cintra is with
the Signal Processing Group,
Departamento de Estat\'{\i}stica,
Universidade Federal de Pernambuco.
E-mail:
\protect\url{rjdsc@stat.ufpe.org}
}
\and
S.~Kulasekera%
\thanks{%
S. Kulasekera and A. Madanayake
are with the
Department of Electrical and Computer Engineering,
University of Akron, Akron, OH, USA,
\protect\url{arjuna@uakron.edu}}
\and
A.~Madanayake${}^\S$
\and
A.~J.~Kozakevicius%
\thanks{%
A.~J.~Kozakevicius
is with the
Departamento de Matem\'atica,
Universidade Federal de Santa Maria,
Santa Maria, RS, Brazil,
\protect\url{alicek@ufsm.br}}
}

\date{}

\newcommand{\myabstract}{%
A low-complexity orthogonal multiplierless
approximation for the 16-point
discrete cosine transform (DCT)
was introduced.
The proposed method
was designed to
possess
a very low computational
cost.
A fast algorithm based on matrix factorization was proposed
requiring only 60~additions.
The proposed architecture
outperforms classical and state-of-the-art algorithms
when assessed as a tool for image and video compression.
Digital VLSI hardware implementations
were also proposed
being physically realized in FPGA technology
and
implemented in 45~nm up to synthesis and place-route levels.
Additionally,
the proposed method was embedded into
a high efficiency video coding (HEVC)
reference software for actual proof-of-concept.
Obtained results show
negligible video degradation
when compared to Chen DCT algorithm in HEVC.
}

\newcommand{\mykeywords}{%
16-point DCT approximation,
Low-complexity transforms,
Image compression,
Video coding
}

\begin{document}

\makeatletter
\if@twocolumn

\twocolumn[%
  \maketitle
  \begin{onecolabstract}
    \myabstract
  \end{onecolabstract}
  \begin{center}
    \small
    \textbf{Keywords}
    \\\medskip
    \mykeywords
  \end{center}
  \bigskip
]
\saythanks

\else

  \maketitle
  \begin{abstract}
    \myabstract
  \end{abstract}
  \begin{center}
    \small
    \textbf{Keywords}
    \\\medskip
    \mykeywords
  \end{center}
  \bigskip
  \onehalfspacing
\fi

\section{Introduction}

The discrete cosine transform (DCT)~\cite{rao1990discrete,britanak2007discrete}
is a pivotal tool for  digital image processing~\cite{Lin2006,bas2010,bas2011}.
Indeed,
the DCT is an important approximation for the
optimal Karhunen-Lo\`eve transform (KLT),
being employed in a multitude of compression standards
due to its remarkable energy compaction properties~\cite{Chang2000, bas2008, bas2011, cb2011, Potluri2013}.
Because of this,
the DCT
has found
at applications
in image and video coding standards,
such as JPEG~\cite{penn1992},
MPEG-1~\cite{roma2007hybrid},
MPEG-2~\cite{mpeg2},
H.261~\cite{h261},
H.263~\cite{h263},
and
H.264~\cite{h264}.
Moreover,
numerous fast algorithms were proposed
for its computation~\cite{arai1988fast,loeffler1991practical,wang1984fast,lee1984new,vetterli1984simple,
hou1987fast,fw1992}.

Designing fast algorithms for the DCT
is a mature area of research~\cite{Heideman1988, loeffler1991practical, Liang2001,  edirisuriya2012vlsi};
thus it is not realistic to expect major advances by means of
standards techniques.
On the other hand,
the development of low-complexity approximations for DCT
is an open field of research.
In particular,
the 8-point DCT was given several approximations,
such as
the signed DCT~\cite{haweel2001new},
the level~1 DCT approximation~\cite{lengwehasatit2004scalable},
the Bouguezel-Ahmad-Swamy (BAS) series of transforms~\cite{
bas2008,
bas2009,
bas2010, bas2011, bas2013},
the rounded DCT (RDCT)~\cite{cb2011}, %
the modified RDCT~\cite{bc2012},
the multiplier-free DCT approximation for RF imaging~\cite{multibeam2012},
and
the improved approximate DCT proposed in~\cite{Potluri2013}.
Such approximations
reduce the computational demands of the DCT evaluation,
leading to low-power consumption
and
high-speed hardware realizations~\cite{Potluri2013}.
At the same time,
approximate transforms can provide adequate numerical
accuracy
for image and video processing.

In
response to the growing need for higher compression rates
related to real time applications~\cite{itu},
the high efficiency video coding (HEVC) video compression format~\cite{hevc1} was proposed.
Different from its predecessors,
the HEVC employs
not only 8$\times$8 size blocks,
but also 4$\times$4, 16$\times$16,
and 32$\times$32.
Several
approximations
for
16-point DCT
based on the integer cosine transform~\cite{Cham1989}
were proposed in~\cite{Cham1991-oreder-16}, \cite{Dong2009} and \cite{fong2012}.
These transformations are derived
from the exact DCT
after scaling the elements of the DCT matrix
and
approximating the resulting real-numbered entries
to integers~\cite{Dong2009}.
Therefore,
real multiplications can be
completely eliminated,
at the expense
of a noticeable increase
in both
the additive complexity
and
the number of required bit-shifting operations~\cite{britanak2007discrete}.

A more restricted class of DCT approximations
prescribe
transformation matrices
with entries defined on the set
$\mathcal{C}=\{0, \pm 1/2, \pm 1, \pm 2\}$.
Because the elements of~$\mathcal{C}$
imply almost null arithmetic complexity,
resulting transformations defined over~$\mathcal{C}$
have \emph{very} low-complexity,
requiring no multiplications
and
a reduced number of bit-shifting operations.
In this context,
methods
providing
16-point low-cost orthogonal transforms
include
the Walsh--Hadamard transform (WHT)~\cite{yarlagadda, valova}, %
BAS-2010~\cite{bas2010},
BAS-2013~\cite{bas2013},
and
the
approximate transform
proposed
in~\cite{bayer201216pt},
here referred to as BCEM approximation.
To the best of our knowledge,
these are the only
16-point DCT approximations defined over $\mathcal{C}$
archived in literature.
Approximations
over $\mathcal{C}$
are
adequate
the HEVC structure~\cite{Potluri2013}
and
are capable of
minimizing
the associated hardware power consumption
as required
by current multimedia market~\cite{itu}.

The aim of this paper is to contribute
to image and video compression
methods
related to
JPEG-like schemes
and
HEVC.
Thus,
we propose
a 16-point approximate DCT,
that requires neither multiplications nor bit-shifting operations.
Additionally,
a fast algorithm is sought,
aiming to minimize the overall computation complexity.
The proposed transform is assessed and compared
with
competing 16-point DCT approximations.
The
realization of the propose DCT approximation
in digital VLSI hardware
as
well as
into a HEVC reference software
is sought.

This paper unfolds as follows.
In Section~\ref{s:proposed},
we propose a new 16-point DCT approximation
and
detail its fast algorithm.
Section~\ref{s:analysis}
presents the performance %
analysis of the introduced transformations
and
compare it to
competing tools
in terms of the computational complexity
coding measures,
and similarity metrics
with respect to the exact DCT.
In Section~\ref{s:image-comp},
a JPEG-like image compression simulation
is described and results are presented.
In Section~\ref{s:hardware},
digital hardware architectures
for the proposed algorithm are supplied
for both \mbox{1-D} and \mbox{2-D} analysis.
A practical real-time video coding scenario is also considered:
the proposed method
is embedded into an open source HEVC standard reference software.
Conclusions and final remarks are given in the last section.

\section{Proposed transform}
\label{s:proposed}

Several
fast algorithm for the DCT
allow
recursive structures,
for which
the computation of
the $N$-point DCT
can be split into the computation
of $\frac{N}{2}$-point
DCT~\cite{blahut,Chen1977,loeffler1991practical,britanak2007discrete,rao1990discrete,yip}.
This is usually the case for algorithms based
on decimation-in-frequency methods~\cite{blahut,Heideman1988}.

In account of the above observation
and judiciously considering permutations and signal changes,
we designed a
16-point transformation
that
splits itself into
two
instantiations
of
the low-complexity matrix
associated to the
8-point RDCT~\cite{cb2011}.
The proposed transformation,
denoted as $\mathbf{T}$,
is given by:
\begin{align*}
\mathbf{T} &  =
  \left[
\begin{rsmallmatrix}\\
1& 1& 1& 1& 1& 1& 1& 1& 1& 1& 1& 1& 1& 1& 1& 1\\
1& 1& 1& 1& 1& 1& 1& 1& -1& -1& -1& -1& -1& -1& -1& -1\\
1& 1& 1& 0& 0& -1& -1& -1& -1& -1& -1& 0& 0& 1& 1& 1\\
1& 1& 0& 0& 0& 0& -1& -1& 1& 1& 0& 0& 0& 0& -1& -1\\
1& 0& 0& -1& -1& 0& 0& 1& 1& 0& 0& -1& -1& 0& 0& 1\\
1& 1& -1& -1& -1& -1& 1& 1& -1& -1& 1& 1& 1& 1& -1& -1\\
1& 0& -1& -1& 1& 1& 0& -1& -1& 0& 1& 1& -1& -1& 0& 1\\
0& 0& -1& 1& 1& -1& -1& 1& -1& 1& 1& -1& -1& 1& 0& 0\\
1& -1& -1& 1& 1& -1& -1& 1& 1& -1& -1& 1& 1& -1& -1& 1\\
1& -1& -1& 1& 0& 0& 1& -1& 1& -1& 0& 0& -1& 1& 1& -1\\
1& -1& 0& 1& -1& 0& 1& -1& -1& 1& 0& -1& 1& 0& -1& 1\\
0& 0& 1& 1& -1& -1& 0& 0& 0& 0& 1& 1& -1& -1& 0& 0\\
0& -1& 1& 0& 0& 1& -1& 0& 0& -1& 1& 0& 0& 1& -1& 0\\
1& -1& 1& -1& 1& -1& 0& 0& 0& 0& 1& -1& 1& -1& 1& -1\\
0& -1& 1& -1& 1& -1& 1& 0& 0& 1& -1& 1& -1& 1& -1& 0\\
1& -1& 0& 0& -1& 1& -1& 1& -1& 1& -1& 1& 0& 0& 1& -1\\
\end{rsmallmatrix}
  \right]
.
\end{align*}
Because the entries of $\mathbf{T}$ are in $\{0,\pm 1\} \subset \mathcal{C}$,
the proposed matrix is a multiplierless operator.
Bit-shifting operations are also unnecessary;
only simple additions are required.
Additionally, the above matrix obeys the condition:
$\mathbf{T} \cdot \mathbf{T}^\top = [\text{diagonal matrix}]$,
where
superscript $\top$ denotes matrix transposition.
Thus,
the necessary conditions for orthogonalizing it
according to the methods described
in~\cite{cintra2011integer}, \cite{cb2011} and \cite{cintra2014low}
are satisfied.
Such procedure
yields
the following
orthogonal 16-point DCT approximation matrix:
\begin{align*}
\mathbf{\hat{C}} =
\mathbf{S} \cdot \mathbf{T},
\end{align*}
where
\begin{align*}
\mathbf{S} = \operatorname{diag}
\Bigg(
&
      \frac{1}{4},\frac{1}{4},
      \frac{1}{\sqrt{12}}, \frac{1}{\sqrt{8}},
      \frac{1}{\sqrt{8}}, \frac{1}{4},
      \frac{1}{\sqrt{12}}, \frac{1}{\sqrt{12}},
\\
&
      \frac{1}{4}, \frac{1}{\sqrt{12}},
      \frac{1}{\sqrt{12}}, \frac{1}{\sqrt{8}},
      \frac{1}{\sqrt{8}}, \frac{1}{\sqrt{12}},
      \frac{1}{\sqrt{12}}, \frac{1}{\sqrt{12}}
\Bigg)
.
\end{align*}

In the context of image and video compression,
the diagonal matrix~$\mathbf{S}$
can be absorbed into
the quantization
step~\cite{bas2008, bas2010, bas2011, cb2011, edirisuriya2012vlsi, tran, cintra2014low}.
Therefore,
under these conditions,
the complexity of the approximation~$\mathbf{\hat{C}}$
can be equated
to the complexity
of the
low-complexity matrix~$\mathbf{T}$~\cite{bayer201216pt, tran}.

Matrix-based fast algorithm design techniques
yield a
sparse matrix factorization
of
$\mathbf{T}$
as given below:
\begin{align*}
 \mathbf{T} = \mathbf{P_2} \cdot %
 \mathbf{M_4} \cdot \mathbf{M_3} \cdot \mathbf{M_2} \cdot \mathbf{P_1} \cdot \mathbf{M_1}\text{,}
\end{align*}
where
\begin{equation}
\label{equation-factorization}
\begin{split}
\mathbf{M}_{1}  & =
  \left[
\begin{rsmallmatrix}\\
\mathbf{I}_8 & \bar{\mathbf{I}}_8 \\
\bar{\mathbf{I}}_8 & -\mathbf{I}_8 \\
\end{rsmallmatrix}
\right]
\text{,}
\quad
\mathbf{P}_{1}
=
\operatorname{diag}\left(
\mathbf{I}_9,
  \left[
\begin{rsmallmatrix}\\
0& 0& 1& 0& 0& 0& 0\\
0& 0& 0& 1& 0& 0& 0\\
0& 0& 0& 0& 0& 0& 1\\
0& 0& 0& 0& 0& 1& 0\\
0& 0& 0& 0& 1& 0& 0\\
0& 1& 0& 0& 0& 0& 0\\
1& 0& 0& 0& 0& 0& 0\\
\end{rsmallmatrix}
  \right]
  \right)
,
\\
\mathbf{M}_{2}  & =  \operatorname{diag}\left(
  \left[
\begin{rsmallmatrix}\\
\mathbf{I}_4 & \bar{\mathbf{I}}_4 \\
\bar{\mathbf{I}}_4 & -\mathbf{I}_4 \\
\end{rsmallmatrix}
\right],
  \left[
\begin{rsmallmatrix}\\
\mathbf{I}_4 & \bar{\mathbf{I}}_4 \\
\bar{\mathbf{I}}_4 & -\mathbf{I}_4 \\
\end{rsmallmatrix}
\right]
\right)
, \\
\mathbf{M}_{3} &=  \operatorname{diag}\left(
  \left[
\begin{rsmallmatrix}\\
1& 0& 0& 1\\
0& 1& \phantom{-}1& 0\\
0& -1& 1& 0\\
1& 0& 0& -1\\
\end{rsmallmatrix}
\right]
,
\left[
\begin{rsmallmatrix}\\
 0&  1&  1& \phantom{-}1\\
-1& -1&  0& 1\\
-1&  1& -1& 0\\
 1&  0& -1& 1\\
\end{rsmallmatrix}
\right]
,
\left[
\begin{rsmallmatrix}\\
 1& 0& \phantom{-}0& \phantom{-}1 \\
 0& 1& 1& 0 \\
 0&-1& 1& 0 \\
-1& 0& 0& 1 \\
\end{rsmallmatrix}
\right]
,
\left[
\begin{rsmallmatrix}\\
 0& \phantom{-}1& \phantom{-}1& \phantom{-}1\\
 1& 1& 0& -1\\
 1& -1& 1& 0\\
 1& 0& -1& 1\\
\end{rsmallmatrix}
\right]
\right)
,
\\
\mathbf{M}_{4}
&=
\operatorname{diag}\left(
  \left[
\begin{rsmallmatrix}\\
1& \phantom{-}1 \\
1& -1\\
\end{rsmallmatrix}
\right]
,
\mathbf{I}_6,
\left[
\begin{rsmallmatrix}\\
1& \phantom{-}1 \\
1& -1\\
\end{rsmallmatrix}
\right],
\mathbf{I}_6
\right)
\end{split}
\end{equation}
and matrix~$\mathbf{P_2}$ performs the simple permutation
(0)(1 8)(2 4 3 11 10 7 12 2)(5 9 13 14 6 5)(15)
in cyclic notation~\cite[p.~77]{Herstein1975}.
Matrices $\mathbf{I}_n$ and $\bar{\mathbf{I}}_n$ denote
the identity and counter-identity matrices of order $n$,
respectively.

\section{Computational Complexity and Evaluation}
\label{s:analysis}

In this section,
we aim at
(i)~assessing
the computational complexity of the proposed approximation,
(ii)~evaluating it in terms
of
approximation error,
and
(iii)~measuring its coding performance~\cite{britanak2007discrete}.
For comparison purposes,
we selected
the following
state-of-the-art
16-point DCT approximations:
BAS-2010~\cite{bas2010},
BAS-2013~\cite{bas2013}
and
the BCEM method~\cite{bayer201216pt}.
We also considered
the classical WHT~\cite{yarlagadda}
and the
exact DCT
as computed according to the Chen DCT algorithm~\cite{Chen1977}.
This latter method is the algorithm
employed in the HEVC codec by~\cite{MiguelCapelo2011}.

\subsection{Arithmetic Complexity}

The computational cost %
of a given transformation is
traditionally measured by its arithmetic complexity,
i.e,
the number of required arithmetic operations
for its computation~\cite{blahut, britanak2007discrete, proakis}.
Considered operations are
multiplications,
additions, and bit-shifting operations~\cite{blahut}.
Table~\ref{T:compl}
lists the operation count
for each arithmetical operation
for all considered methods.
Total operation count is also provided.

\begin{table}[]
\caption{Arithmetic complexity assessment}
\label{T:compl} %
\centering
\begin{tabular}{l c c c c }
\toprule
\multirow{2}{*}{Transform} &
\multicolumn{4}{c}{Operation count}
\\
\cmidrule{2-5}
&  Multiplication &  Addition   &   Bit-shifting  &  Total  \\
\midrule
Chen DCT~\cite{Chen1977}
            &  44 &  74   &  0      &  118 \\
WHT         &  0  &  64   &  0      &  64 \\
BAS-2010    &  0  &  64   &  8       & 72\\
BAS-2013    &  0  &  64   &  0   & 64\\
BCEM
            &  0  &  72   &  0      &  72 \\
Proposed    &  0  &  \bf{60}   &  0      &  \bf{60}\\
\bottomrule
\end{tabular}
\end{table}

The proposed transform
showed
6{.}25\% less total operation count
when compared with the WHT or BAS-2013 approximation.
Considering BAS-2010 or
the BCEM approximation, %
the introduced approximation
required
16{.}67\% less operation overall.
As a more strict complexity assessment,
even if we take  only the additive complexity into account,
the proposed transformation can still outperform all considered methods.
It is also noteworthy that
the proposed method
has the lowest multiplicative complexity
among all considered methods.
Moreover,
to the best of our knowledge,
the proposed transformation
outperforms
any
meaningful 16-point DCT approximation archived in literature.

\subsection{Similarity Measures}

For the approximation error analysis,
we considered
three
tools:
the DCT distortion~\cite{wien},
the
total error energy~\cite{cb2011},
and
the mean square error~(MSE)~\cite{britanak2007discrete,rao1990discrete}.
This set of measures
determines
the similarity
between the exact DCT matrix and
a
given approximation.
These quality metrics are briefly described as follows.

Let $\mathbf{C}$ be the exact $N$-point DCT matrix
and
$\mathbf{\tilde{C}}$ be a given $N$-point DCT approximation.
Adopting the notation employed in~\cite{fong2012},
the DCT distortion of $\mathbf{\tilde{C}}$
is given by: %
\begin{align*}
\text{$d_{2}(\mathbf{\tilde{C}})$}
=
1
-
\frac{1}{N}
\cdot
\left\|
\operatorname{diag}
\left(
\mathbf{C}\cdot\mathbf{\tilde{C}}^\top
\right)
\right\|^2
,
\end{align*}
where
$\left \| \ \cdot \ \right \|$
is the Euclidean norm~\cite{Watkins2004}.
The DCT distortion captures
the difference
between
the exact DCT matrix
and a candidate approximation
by quantifying the orthogonality among the basis vectors
of both transforms~\cite{fong2012}.
Taking the basis vectors of the exact DCT
and
a given approximation as filter coefficients,
the total error energy~\cite{cb2011}, %
measures the spectral proximity
between the corresponding transfer functions~\cite{haweel2001new}.
Invoking Parseval theorem~\cite{Oppenheim2006},
the total error energy can be evaluated according to:
\begin{align*}
\epsilon(\mathbf{\tilde{C}})
=
\pi \cdot
\left \|
\mathbf{C} - \mathbf{\tilde{C}}
\right \|^2_\text{F}
,
\end{align*}
where $\| \cdot \|_\text{F}$
is the Frobenius norm~\cite{Watkins2004}.%

The MSE is a well-established proximity measure~\cite{britanak2007discrete}.
The MSE between
$\mathbf{C}$
and
$\mathbf{\tilde{C}}$
is given by~\cite{Liang2001,britanak2007discrete}:
\begin{align*}
\operatorname{MSE}(\mathbf{\tilde{C}})
=
\frac{1}{N}
\cdot
\operatorname{tr}
\left (
(\mathbf{C}-\mathbf{\tilde{C}})
\cdot
\mathbf{R}
\cdot
(\mathbf{C} - \mathbf{\tilde{C}})^\top
\right )
,
\end{align*}
where
$\operatorname{tr}{(\ \cdot \ )}$ is the trace function~\cite{seber07}
and
$\mathbf{R}$ is the covariance matrix
of the input signal.
Assuming the first-order stationary Markov process model
for the input data,
we have that
$\mathbf{R}_{[i,j]} = \rho^{|i-j|}$,
for $i,j = 1,2,\ldots,N$,
and
the correlation coefficient $\rho$ is
set to 0{.}95~\cite{Liang2001, britanak2007discrete}.
This particular model is suitable
for real signals and natural
images~\cite{britanak2007discrete, Liang2001, cintra2014low}.
The minimization of MSE values
indicates proximity to the exact DCT~\cite{britanak2007discrete}.

\subsection{Coding Measures}

We adopted two coding measures:
the transform coding gain~\cite{fong2012}
and
the
transform efficiency~\cite{britanak2007discrete}.
The transform coding gain
quantifies the coding
or data compression performance of
an orthogonal transform~\cite{fong2012,britanak2007discrete}.
This measure is given by~\cite{britanak2007discrete,malvar}:
\begin{align*}
C_{g}(\mathbf{\tilde{C}})
=
10
\cdot
\log_{10}
\left\{
\frac{\frac{1}{N}\sum_{i=0}^{N-1} s_{ii}}
{
\left[
\prod_{i=0}^{N-1}
\left(
s_{ii}
\cdot
\sqrt{\sum_{j=0}^{N-1}\mathbf{\tilde{c}}_{ij}^2}
\right)
\right]^\frac{1}{N}
}
\right\}
,
\end{align*}
where
$s_{ij}$ and $\mathbf{\tilde{c}}_{ij}$ are the $(i,j)$-th entry
of
$\mathbf{\tilde{C}} \cdot \mathbf{R} \cdot \mathbf{\tilde{C}}^\top$ and $\mathbf{\tilde{C}}$, respectively.

On the other hand,
the transform efficiency~\cite{britanak2007discrete}
is an alternative method to compute the compression
performance.
Denoted by $\eta$,
the transform efficiency is furnished by:
\begin{align*}
\eta(\mathbf{\tilde{C}})
=
\frac{\sum_{i=0}^{N-1}\left|s_{ii}\right|}
{\sum_{i=0}^{N-1}\sum_{j=0}^{N-1}\left|s_{ij}\right|}
\times
100
.
\end{align*}
Quantity
$\text{$\eta(\mathbf{\tilde{C}})$}$
indicates the data decorrelation capability
of the transformation. %
The KLT achieves optimality with respect to this measure,
presenting a transform efficiency of 100~\cite{britanak2007discrete}.

\subsection{Results}

Table~\ref{T:perf}
summarizes
the results for the
above detailed
similarity and coding measures.
For each figure of merit,
we emphasize in bold
the two best
measurements.
The proposed transform
displays
consistently
good performance
according to all considered criteria.
This fact contrasts with existing transformations,
which tend to excel in terms of similarity measures,
but perform limitedly in terms of coding performance;
and vice-versa.
Therefore,
the proposed transform
offers a compromise,
while still
achieving state-of-the-art performance.

\begin{table}[]
\caption{Performance analysis}
\label{T:perf} %
\centering
\begin{tabular}{l c c c c c }
\toprule
\multirow{4}{*}{Transform}
&
\multicolumn{5}{c}{Measures}
\\
\cmidrule{2-6}
&
\multicolumn{3}{c}{Approximation} &
\multicolumn{2}{c}{Coding}
\\
\cmidrule{2-4}
\cmidrule{4-6}
        &  $d_2$ &  $\epsilon$ &  MSE &  $C_g$ &  $\eta$ \\
\midrule
DCT     &
0       &  0        &  0       &   9.4555  &  88.4518     \\
WHT     &
0.8783  &  92.5631  &  0.4284  &   8.1941  &  70.6465     \\
BAS-2010 &
0.6666  &  64.749   &   0.1866 &  \textbf{8.5208}   &  \textbf{73.6345}   \\
BAS-2013  &
0.5108   &  54.6207      &  0.132   &  8.1941   &  70.6465  \\
BCEM &
\textbf{0.1519}  &  \textbf{8.0806}   &  \textbf{0.0465}   &  7.8401   &  65.2789    \\
Proposed   &
\textbf{0.3405}  &  \textbf{30.323}   &  \textbf{0.0639}   &   \textbf{8.295}  &  \textbf{70.8315}
\\
\bottomrule
\end{tabular}
\end{table}

\section{Application to image compression}
\label{s:image-comp}

The proposed approximation
was submitted to the image compression simulation
methodology
originally
introduced in~\cite{haweel2001new}
and employed
in~\cite{bas2008,bas2010,bas2011,bas2013,cb2011,bayer201216pt}.
In our experiments a set of 45 512$\times$512
8-bit grayscale images obtained from
a standard public image bank~\cite{uscsipi}
was
considered to validate the proposed algorithm.
We adapted the JPEG-like compression scheme
for the 16$\times$16 matrix case,
as suggested
in~\cite{bas2010}.

The adopted image compression method is detailed as follows.
An input 512$\times$512 image was divided into 16$\times$16
disjoint blocks $\mathbf{A}_k$,
$k=0,1,\ldots,31$.
Each block $\mathbf{A}_k$
was
\mbox{2-D} transformed
according to
$\mathbf{B}_k = \mathbf{\tilde{C}}\cdot\mathbf{A}_k\cdot\mathbf{\tilde{C}}^\top$,
where
$\mathbf{B}_k$ is a frequency domain image block
and
$\mathbf{\tilde{C}}$ is a given approximation matrix.
Matrix $\mathbf{B}_k$ contains
the 256~transform domain coefficients
for each block.
Adapting the zigzag sequence~\cite{pao1998}
for the 16$\times$16 case,
we retained only the $r$ initial coefficients
and
set the remaining coefficients to zero~\cite{
haweel2001new,
cb2011,
bayer201216pt,
bas2008,
bas2010,
bas2011,
bas2013},
generating $\mathbf{B'}_k$.
Subsequently,
the inverse transformation was applied to
$\mathbf{B'}_k$
according to:
$\mathbf{A'}_k =
\mathbf{\tilde{C}}^\top\cdot\mathbf{B'}_k\cdot\mathbf{\tilde{C}}$.
The above procedure was repeated for each block.
The rearrangement of all blocks $\mathbf{A'}_k$
reconstructs the image,
which can be assessed for quality.

Image degradation
was
evaluated using two different quality measures:
(i)~the peak signal-to-noise ratio (PSNR)
and
(ii)~the structural similarity index
(SSIM)~\cite{Wang2004}---a generalization of
the universal image quality index~\cite{wang2002universal}.
In contrast to the PSNR,
SSIM definition
takes advantage of known characteristics
of the human visual system~\cite{Wang2004}.
Following the methodology adopted in~\cite{cb2011} and \cite{bayer201216pt},
we calculated average PSNR and SSIM values for all 45~images.

\begin{figure}[]
\centering
 \subfigure[Average PSNR]{\includegraphics[width=0.45\textwidth]{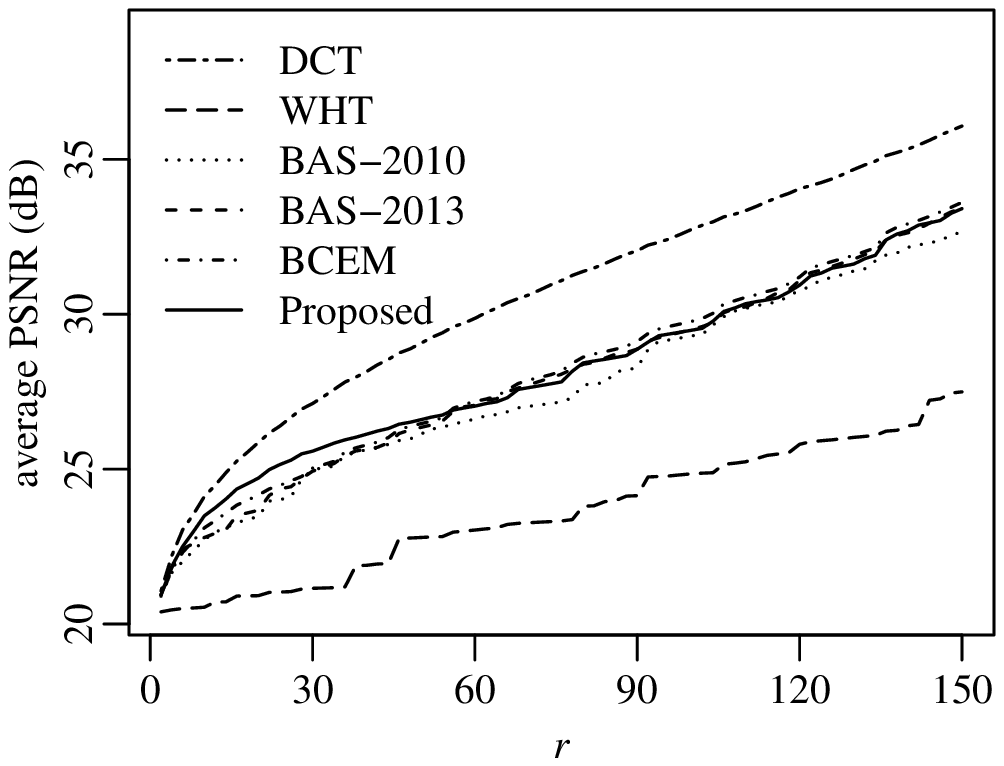}
 \label{f:psnr}}
 \subfigure[Absolute percentage error of PSNR]{\includegraphics[width=0.45\textwidth]{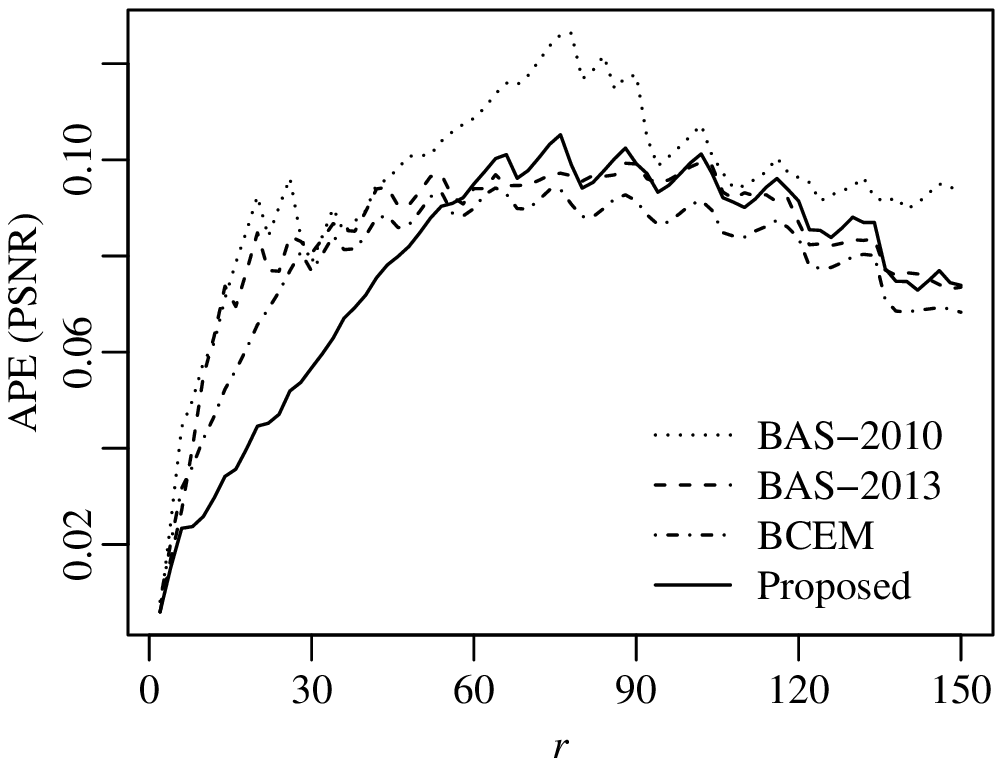}
 \label{f:apepsnr}}
 \caption{PSNR results for all considered transforms under several compression rates}
\label{f:epsnr}
\end{figure}

\begin{figure}[]
\centering
 \subfigure[Average SSIM]{\includegraphics[width=0.45\textwidth]{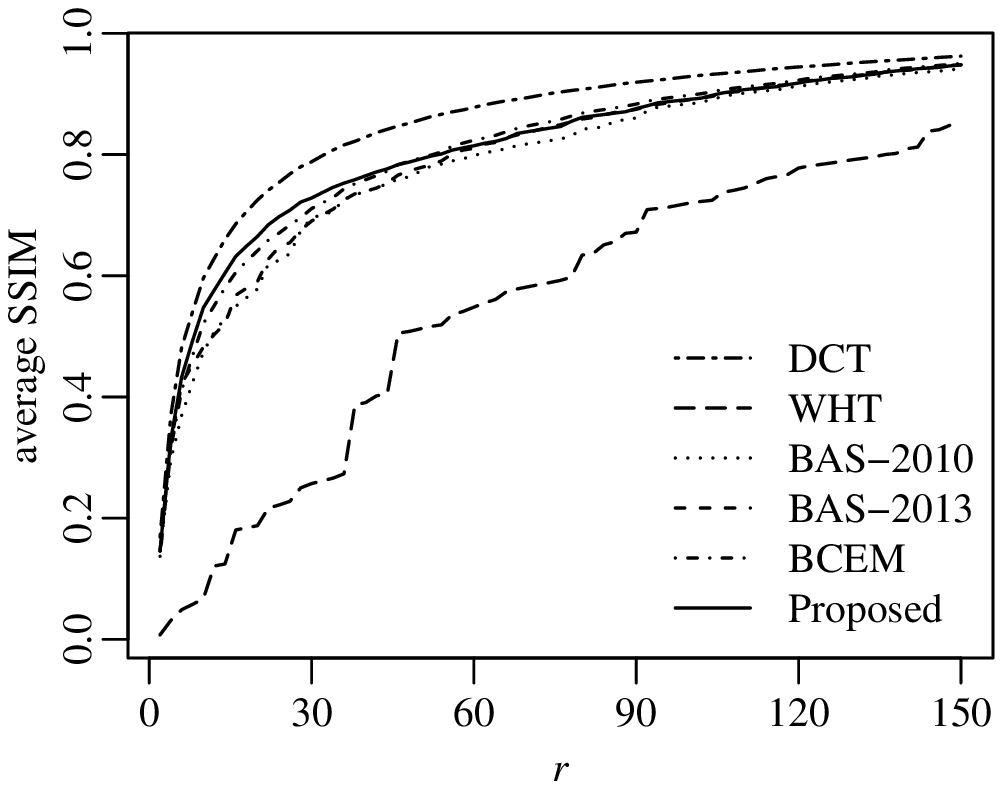}
 \label{f:mssim}}
\subfigure[Absolute percentage error of SSIM]{\includegraphics[width=0.45\textwidth]{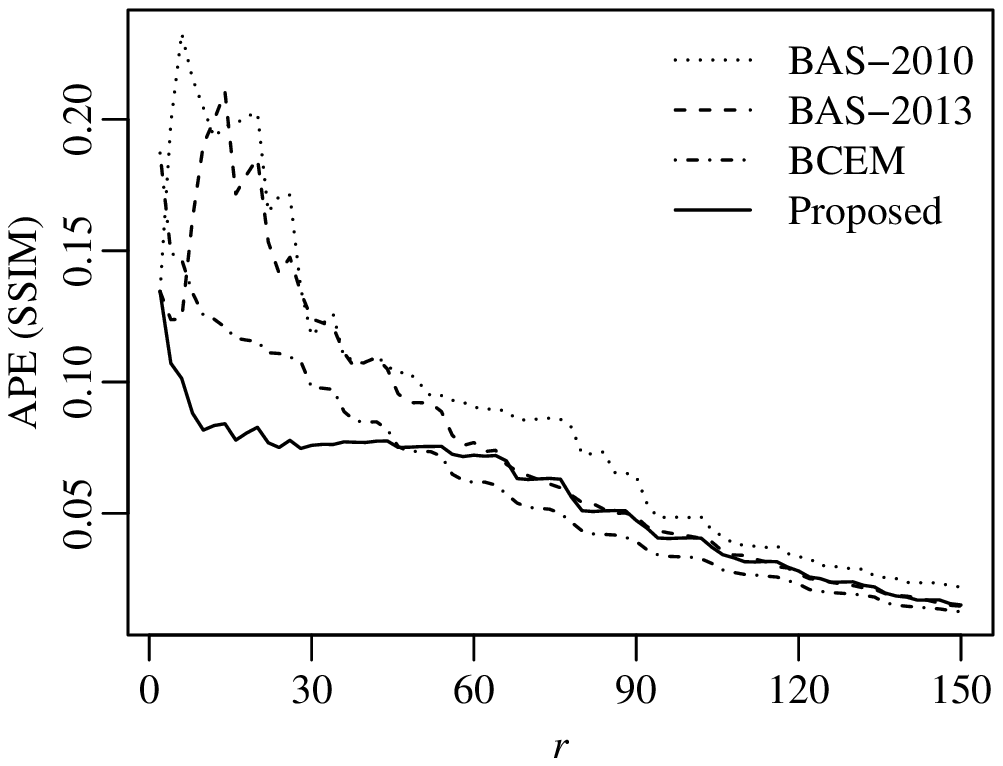}
 \label{f:apemssim}}
 \caption{SSIM results for all considered transforms under several compression rates}
 \label{f:emssim}
\end{figure}

Fig.~\ref{f:psnr} and Fig.~\ref{f:mssim}
show average PSNR and SSIM measurements,
respectively.
Additionally,
we considered absolute percentage error~(APE)
measurements of PSNR and SSIM
with respect to the exact DCT.
Results are displayed
in Fig.~\ref{f:apepsnr} and Fig.~\ref{f:apemssim},
for PNSR and SSIM, respectively.
APE figures for the WHT
are absent because
their values were exceedingly high,
being located outside of the plot range.

According to Figs.~\ref{f:epsnr} and \ref{f:emssim},
the proposed transform outperforms other methods
for $r \leq 50$,
which correspond to high-compression rates.
Therefore,
the proposed transform is
in consonance with
ITU recommendation for high-compression coding
in real time applications~\cite{itu}.
For $r>50$,
discussed methods are essentially comparable
in terms of image degradation.

\begin{figure}[]
\centering
\subfigure[DCT \scriptsize{($\text{PSNR}=28.55$)}]
{\includegraphics[scale=0.21]{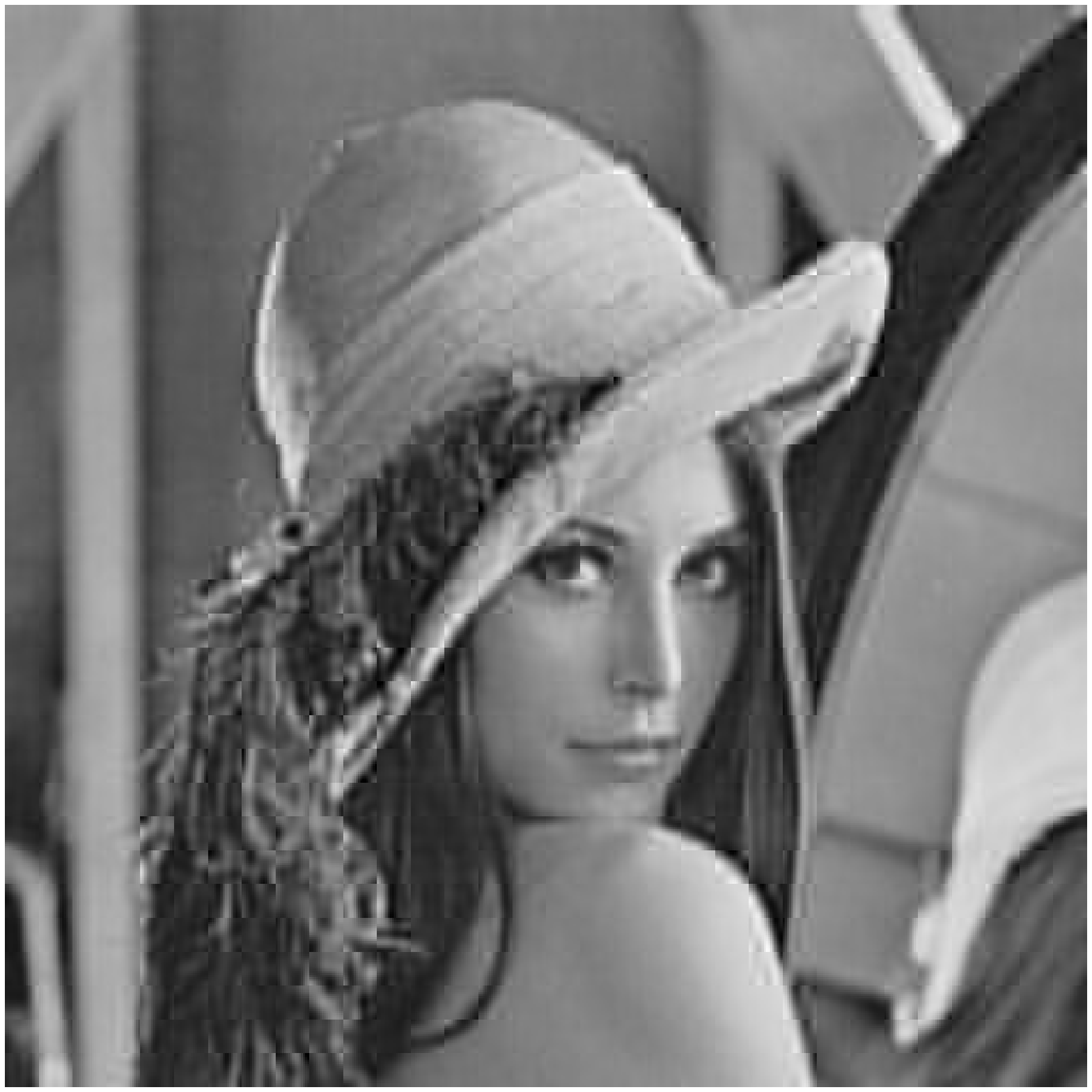}
\label{f:lenadct}}\
\subfigure[WHT \scriptsize{($\text{PSNR}=21.20$)}]
{\includegraphics[scale=0.21]{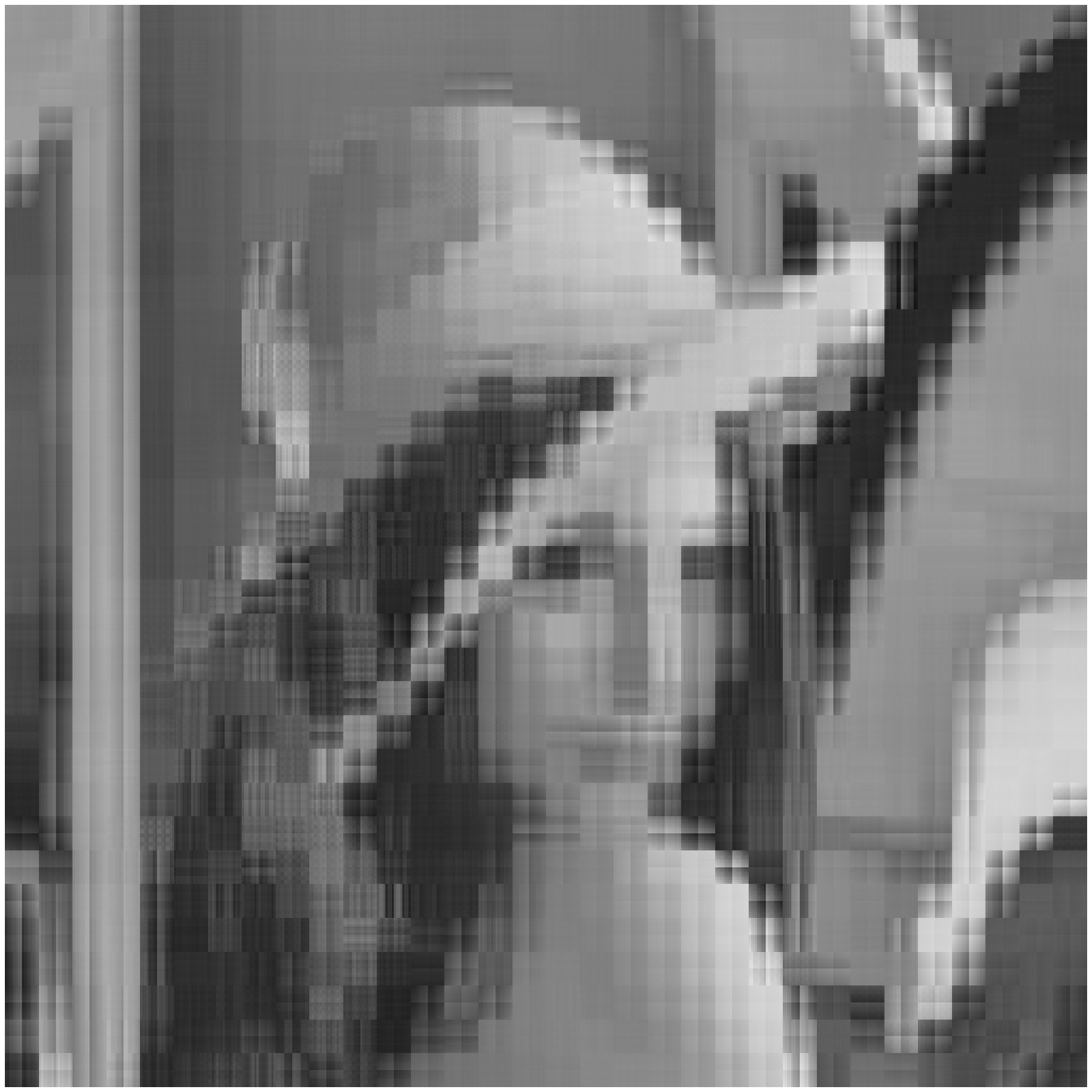}
\label{f:lenawht}}\
\subfigure[BAS-2010 \scriptsize{($\text{PSNR}=25.27$)}]
{\includegraphics[scale=0.21]{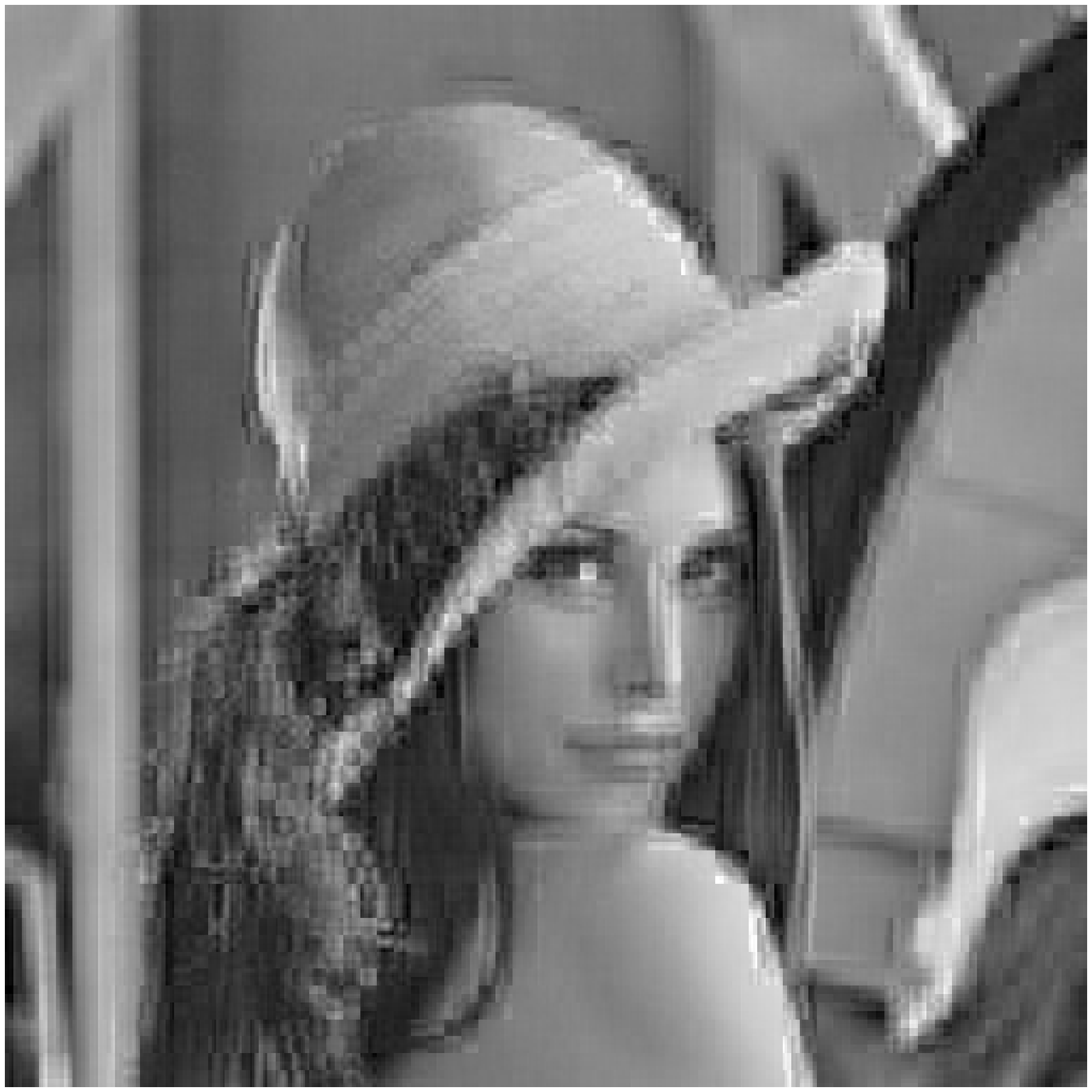}
\label{f:lenabas10}}\\
\subfigure[BAS-2013 \scriptsize{($\text{PSNR}=25.79$)}]
{\includegraphics[scale=0.21]{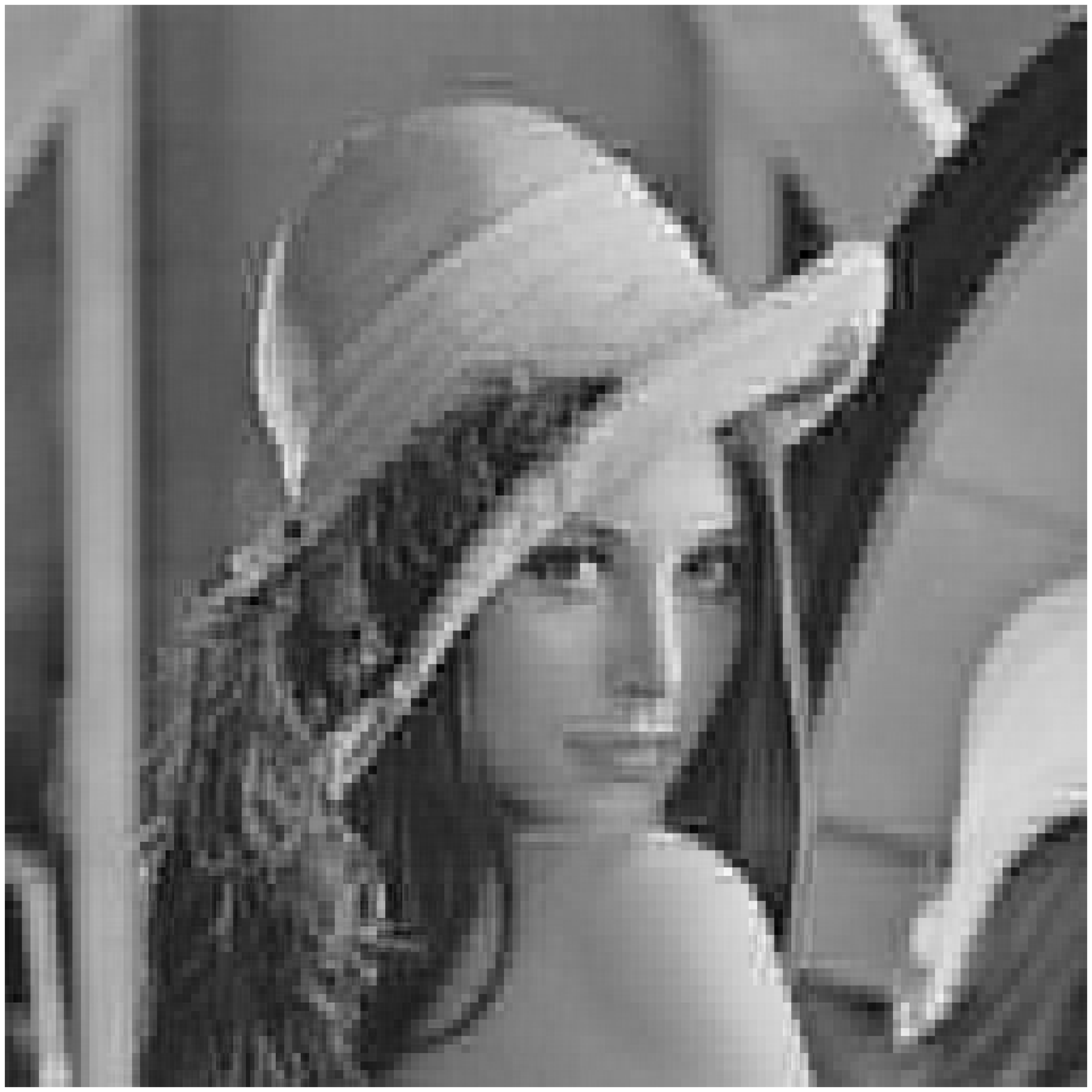}
\label{f:lenabas13}}\
\subfigure[BCEM \scriptsize{($\text{PSNR}=25.75$)}]
{\includegraphics[scale=0.21]{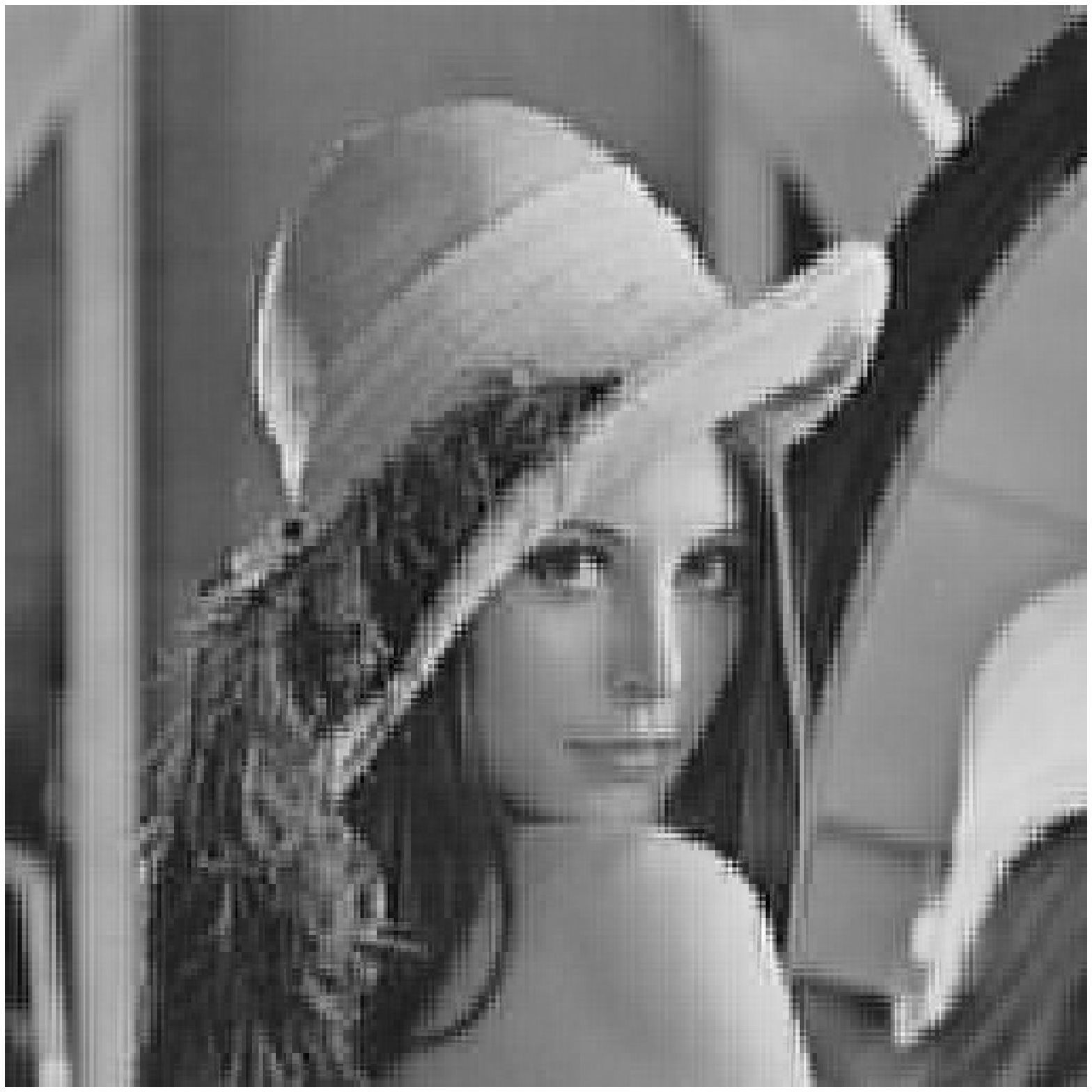}
\label{f:lenabayer12}}\
\subfigure[Proposed \scriptsize{($\text{PSNR}=27.13$)}]
{\includegraphics[scale=0.21]{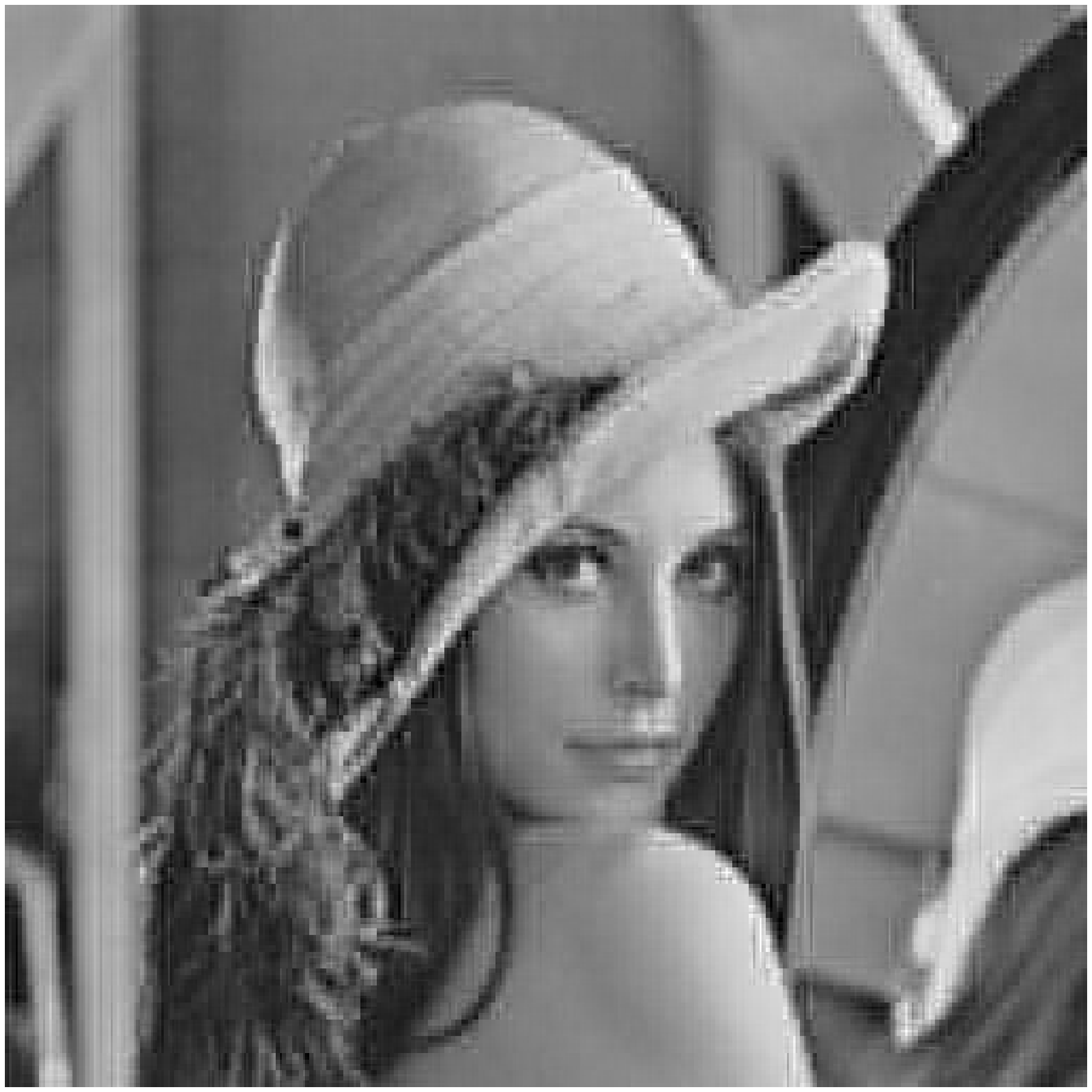}
\label{f:lenaprop}}

\caption{Compressed \textit{Lena} image using all considered transforms,
for $r = 16$}
\label{f:lenna}
\end{figure}

As a qualitative comparison,
Fig.~\ref{f:lenna}
shows
the compressed \textit{Lena} image
at $r=16$ (93.75\% compression)
obtained from each considered method.
The proposed transform offered less pixelation and block artifacts;
demonstrating its adequacy
for high-compression rate scenarios.

\section{Digital Architecture and Realization}
\label{s:hardware}

In this section,
hardware architectures
for the
proposed
16-point approximate DCT
are detailed.
Both \mbox{1-D} and \mbox{2-D} transformations are addressed.
Introduced architectures were submitted to
(i)~Xilinx
field programmable gate array (FPGA)
implementations and
(ii)~CMOS 45~nm
application specific integrated circuit (ASIC)
implementation up to the synthesis level.
Additionally,
in order to assess
the performance of the proposed algorithm in real time video coding,
the introduced approximation was also embedded
into an HEVC reference software~\cite{refsoft}.

\subsection{Architecture for the 16-point DCT approximation}

\begin{figure}
\centering
\scalebox{0.70}{\input{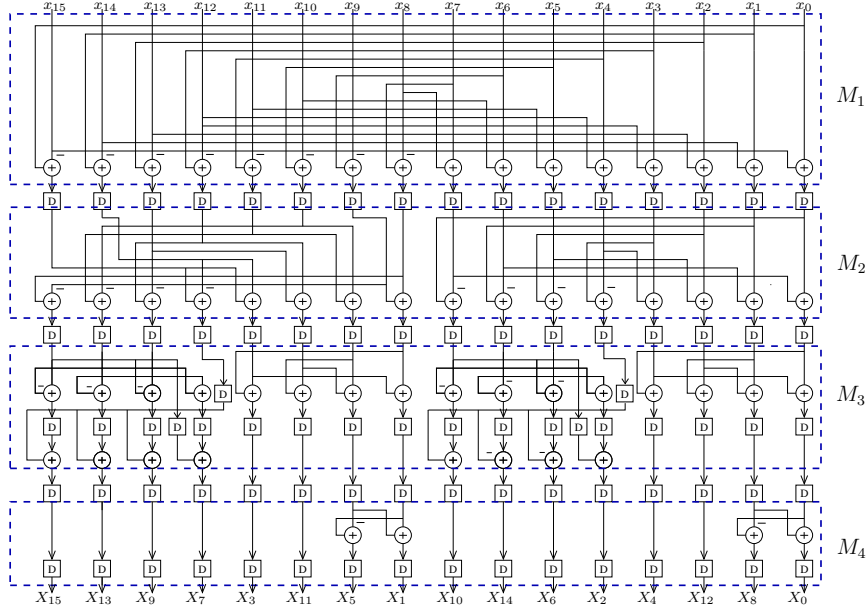}}
\caption{Architecture of the proposed 16-point DCT approximation}\label{fig:16arch}
\end{figure}

\begin{figure}
\centering
\scalebox{1.25}{\input{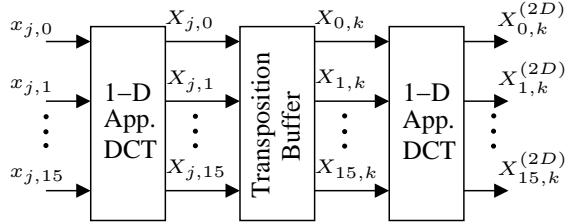}}
\caption{%
Two-dimensional approximate transform
by means of \mbox{1-D} approximate transform.
Signal $x_{k,0},x_{k,1},\ldots$,
corresponds to the rows of the input image;
$X_{k,0},X_{k,1},\ldots$ indicates the transformed rows;
$X_{0,j},X_{1,j},\ldots$ indicates the columns of
the transposed row-wise transformed image;
and
$X_{0,j}^\text{(2-D)},X_{1,j}^\text{(2-D)},\ldots$
indicates the columns of
the final \mbox{2-D} transformed image}
\label{fig:2darch}
\end{figure}

The \mbox{2-D} version of the 16-point DCT approximation architecture
was realized using two \mbox{1-D} transforms and a transpose buffer.
This is possible because the proposed approximation
inherits the separable kernel property of
the exact DCT~\cite{Smith1997}. %
The first instantiation of the approximate DCT block
furnishes
a row-wise transform computation of the input image,
while the second implementation furnishes a column-wise
transformation of the previous intermediate result.
A real time row-parallel transposition buffer circuit
is required
in between
the \mbox{1-D} transformation blocks.
Such block ensures data ordering
for converting the row-transformed data
from the first DCT approximation circuit
to a transposed format
as required by the second DCT approximation circuit.
Both \mbox{1-D} transformation blocks and the transposition buffer
were
initially modeled and tested in Matlab Simulink;
then they were combined to furnish the complete
\mbox{2-D} approximate transform.
Fig.~\ref{fig:16arch} depicts the architecture
for the proposed \mbox{1-D} approximate DCT.
We emphasize in dashed boxes
the blocks
$M_1$, $M_2$, $M_4$, and $M_4$,
which
correspond to the realization
of
sparse
matrices~$\mathbf{M_1}$,
$\mathbf{M_2}$,
$\mathbf{M_3}$,
and
$\mathbf{M_4}$,
respectively,
as shown in the equation set~\eqref{equation-factorization}.
Fig.~\ref{fig:2darch} shows the implementation
of the \mbox{2-D} transform
by means of the \mbox{1-D} transforms.

\subsection{FPGA and ASIC realizations and results}

The above discussed architecture was physically realized
on a FPGA based rapid prototyping system
for various register sizes and
tested using on-chip hardware-in-the-loop co-simulation.
The architecture was designed for digital realization within the
MATLAB environment using the Xilinx System Generator.
Xilinx Virtex-6 XC6VLX240T-1FFG1156 device
was employed
to physically realize the architecture
on FPGA with fine-grain pipelining for increased throughput.
The realization was verified on FPGA chip
using a Xilinx ML605 board at a clock frequency of 50~MHz.
The FPGA realization was tested with 10{,}000 random
16-point input test vectors using hardware co-simulation.
Test vectors were generated from
within the MATLAB environment
and
routed to the physical FPGA device using JTAG based hardware co-simulation.
Then
measured data from the FPGA was routed back
to MATLAB memory space.

Evaluation of hardware complexity and real time performance
considered the following metrics:
the number of used configurable logic blocks (CLB),
flip-flop (FF) count,
critical path delay ($T_\text{cpd}$),
and
the maximum operating frequency ($F_{\text{max}}$) in~MHz.
The \texttt{xflow.results} report file was accessed to obtain the above results.
Dynamic ($D_p$)
and
static power ($Q_p$) consumptions
were estimated using the Xilinx XPower Analyzer.
Results are shown in Table~\ref{table16}.

For the ASIC implementation,
the hardware description
language code was ported to 45~nm CMOS technology and subject to
synthesis and place-and-route steps using the Cadence Encounter.
The FreePDK,
a free open-source ASIC standard cell library
at the 45~nm node,
was used for this purpose.
The supply voltage of the CMOS realization was fixed at
$V_\text{DD} = 1.1~\mathrm{V}$
during estimation of power consumption and logic delay.
The adopted figures of merit for the ASIC synthesis
were:
area ($A$) in~$\mathrm{mm^2}$,
area-time complexity ($AT$) in $\mathrm{mm}^2 \cdot \mathrm{ns}$,
area-time-squared complexity ($AT^2$) in $\mathrm{mm}^2 \cdot \mathrm{ns}^2$,
dynamic ($D_p$) power in ($mW/MHz$) and static ($Q_p$) power consumption in watts,
critical path delay ($T_{cpd}$) in~ns,
and
maximum operating frequency ($F_{\text{max}}$) in MHz.
Results are displayed in Table~\ref{asic16}.

\begin{table}%
\centering
\caption{Hardware resource consumption
and power consumption
for the proposed 2-D 16-point DCT approximation}
\label{table16}
\begin{tabular}{l c c c c c} %
\toprule
\parbox{0.7cm}{\centering CLB} &
\parbox{0.3cm}{\centering FF } &
$T_\text{cpd}$ ($\mathrm{ns}$) &
$F_{\text{max}}$ ($\mathrm{MHz}$) &
$D_p$ ($\mathrm{mW/MHz}$) &
$Q_p$ ($\mathrm{W}$)
\\
\midrule
1408 & 4600 & 3.7 & 270.27 & 6.91 & 3.481 \\
\bottomrule
\end{tabular}
\end{table}

\begin{table}%
\centering
\caption{Hardware resource consumption
for CMOS 45nm ASIC place-route implementation of
the proposed 2-D 16-point DCT approximation}
\label{asic16}
\begin{tabular}{c c c c c c c} %
\toprule
Area\tiny($\mathrm{mm}^2$) &
$AT$ &
$AT^2$  &
$T_\text{cpd}$($\mathrm{ns}$) &
$F_{\text{max}}$($\mathrm{MHz}$) &
$D_p$($\mathrm{mW/MHz}$) &
$Q_p$($\mathrm{mW}$)\\
\midrule
0.585 & 4.896 & 40.98 & 8.37 & 119.47 & 0.311 & 216.2  \\
\bottomrule
\end{tabular}
\end{table}

Among the considered competitors,
the BAS-2010~\cite{bas2010} showed
arithmetic complexity and coding performance
similar to
the proposed transform.
For comparison purposes the \mbox{1-D}
versions of the BAS-2010 approximation
and the proposed 16-point approximation were realized
on a Xilinx Virtex-6 XC6VLX240T-1FFG1156 device
as well as were ported to 45~nm CMOS technology and subject to
synthesis and place-and-route steps using the Cadence Encounter.
The results are shown in Table~\ref{FPGAresults} and Table~\ref{ASICResults}.
Compared to the BAS-2010, the proposed transform is faster when both the FPGA implementation and CMOS synthesis is considered while having
similar performance in hardware usage and dynamic power consumption. Importantly, the proposed is better in image quality as evidenced
by Fig.~\ref{f:lenna}.

\begin{table}%
\begin{center}
\caption{Hardware resource consumption of the 1-D approximations using Xilinx Virtex-6 XC6VLX240T-1FFG1156 device}
\label{FPGAresults}
\begin{tabular}{l c c c c c c} %
\toprule
\parbox{2cm}{\centering Transform} &
 \parbox{0.5cm}{\centering CLB} &
\parbox{0.5cm}{\centering FF } &
$T_\text{cpd}$ ($\mathrm{ns}$) &
$F_{\text{max}}$ ($\mathrm{MHz}$) &
$D_p$ ($\mathrm{mW/MHz}$) &
$Q_p$ ($\mathrm{W}$)\\ \midrule
  {\centering BAS-2010} & 430 & 1440 & 1.950 & 512.82 & 4.54 & 3.49 \\ %
 {\centering Proposed} & 421 & 1372 & 1.900 & 526.31 & 4.22 & 3.49 \\
 \bottomrule
\end{tabular}
\end{center}
\end{table}

\begin{table}%
\begin{center}
\caption{Hardware resource consumption for CMOS 45nm ASIC place-route implementation of the 1-D approximations}
\label{ASICResults}
\begin{tabular}{l c c c c c c c } %
\toprule
Transform &
Area\tiny($\mathrm{mm}^2$) &
$AT$ &
$AT^2$  &
$T_\text{cpd}$\tiny($\mathrm{ns}$) &
$F_{\text{max}}$\tiny($\mathrm{MHz}$) &
$D_p$\tiny($\mathrm{mW/MHz}$) &
$Q_p$\tiny($\mathrm{mW}$) \\ \midrule
 {\centering BAS-2010} & 0.169 & 0.843 & 4.21 & 4.994 & 200.24 & 0.093 & 70.47 \\ %
 {\centering Proposed} & 0.183 & 0.895 & 4.38 & 4.895 & 204.29 & 0.095 & 78.73 \\
 \bottomrule
\end{tabular}
\end{center}
\end{table}

\subsection{Real time video compression software implementation}

In order to assess
real-time video coding
performance,
the proposed  approximation
was embedded into
the open source HEVC standard reference software
by the Fraunhofer Heinrich Hertz Institute~\cite{refsoft}.
The original transform prescribed in the selected HEVC reference software
is
the scaled approximation
of Chen DCT algorithm~\cite{Chen1977,MiguelCapelo2011}
and the software can process image block sizes of
4$\times$4,
8$\times$8,
16$\times$16,
and
32$\times$32.

\begin{figure}
\centering
\scalebox{1}{\input{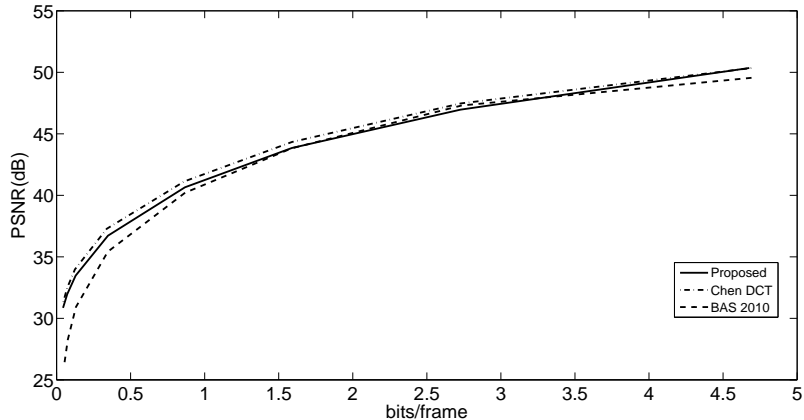}}
\caption{RD curves for `BasketballPass' test sequence}
\label{fig:RD}
\end{figure}

\begin{figure}
\centering
\subfigure[Chen DCT ($\text{QP}=0$)]
{\includegraphics[width=0.47\linewidth]{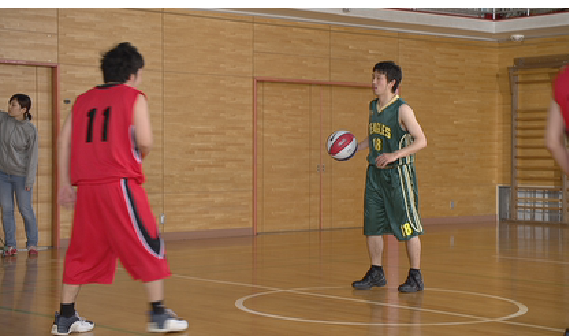}}
 \label{Chen QP0}
\subfigure[Proposed DCT ($\text{QP}=0$)]
{\includegraphics[width=0.47\linewidth]{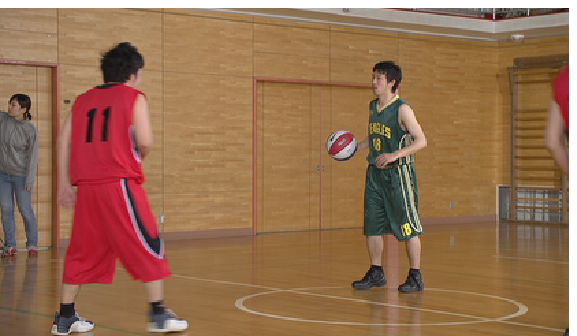}}
 \label{Proposed QP0}
\\*
\subfigure[Chen DCT ($\text{QP}=32$)]
{\includegraphics[width=0.47\linewidth]{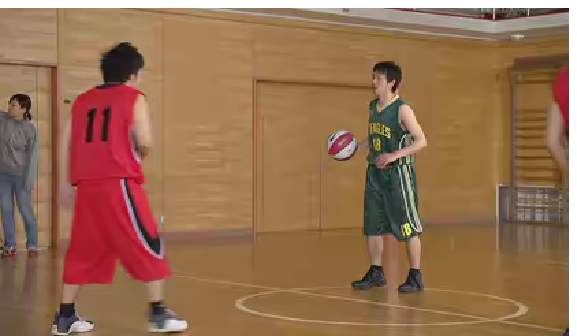}}
 \label{Chen QP32}
\subfigure[Proposed DCT ($\text{QP}=32$)]
{\includegraphics[width=0.47\linewidth]{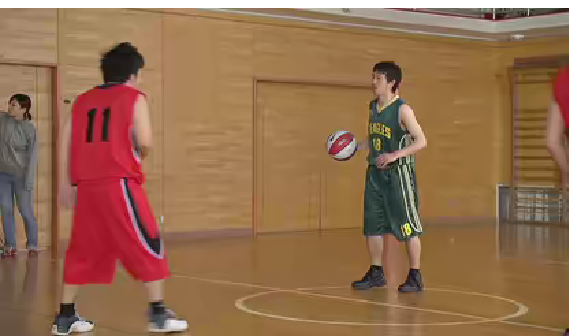}}
 \label{Proposed QP32}
\\*
 \subfigure[Chen DCT ($\text{QP}=50$)]
{\includegraphics[width=0.47\linewidth]{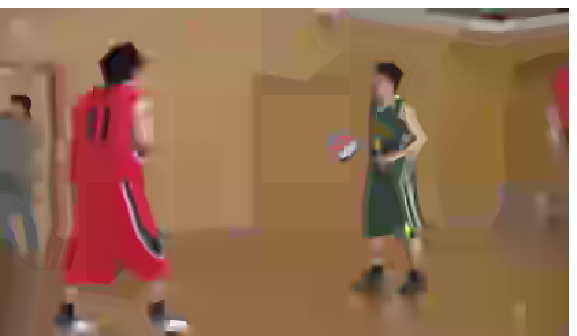}}
 \label{Chen QP50}
\subfigure[Proposed DCT ($\text{QP}=50$)]
{\includegraphics[width=0.47\linewidth]{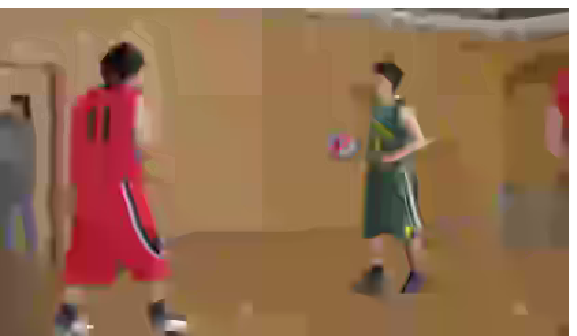}}
 \label{Proposed QP50}

 \caption{Selected frames from `BasketballPass' test video coded
by means of the Chen DCT and the proposed 16-point DCT approximation
for
$\text{QP}= 0$ (a--b),
$\text{QP} = 32$ (c--d),
and
$\text{QP} = 50$ (e--f)}
\label{F:frames}
\end{figure}

Our methodology
consists of replacing the 16$\times$16 DCT algorithm of
the reference software by the proposed 16-point approximate algorithm.
Algorithms were evaluated for their effect on
the overall performance of the encoding process.
For such,
we obtained
rate-distortion~(RD) curves for standard video sequences~\cite{rdcurve}.
The quantization point~(QP)
varied from~0 to~50 to obtain the curves
and
the resulting PSNR values
with the bit rate values measured in bits per frame
were recorded for the proposed algorithm,
the Chen DCT algorithm,
and the BAS-2010~\cite{bas2010} algorithm.
Fig.~\ref{fig:RD} depicts the obtained RD curves
for the `BasketballPass' test sequence.
Fig.~\ref{F:frames} shows particular
416$\times$240 frames
for the test video sequence `BasketballPass'
with
$\text{QP} \in \{0, 32, 50\}$.

The RD curves and selected frames
reveal that the difference between the original HEVC and
the implementation with the proposed approximation is negligible.
In fact, in Fig.~\ref{fig:RD} the maximum PSNR difference is 0.56~dB,
which is very low.
Fig.~\ref{F:frames} shows that both encoded video
streams are almost identical.
These results confirm the adequacy of the proposed scheme.

\section{Conclusion}

This work proposed a new orthogonal 16-point DCT approximation.
The introduced transform
offers
a very low computational cost,
outperforming---to the best of our knowledge---all competing
methods.
Moreover,
the proposed transform
performed well as an image compression tool,
specially at high compression rate scenarios.
By means of
(i)~comprehensive computational simulations,
(ii)~hardware implementation (both in FPGA and ASIC),
and
(iii)~software embedding,
we demonstrated
the adequacy
and
efficiency
of the proposed method,
which is suitable for codec schemes,
like the HEVC.
Additionally,
the introduced transformation offers an unusual good performance balance
among several metrics, as shown in Table~\ref{T:perf}.
This suggests that the applicability
of proposed transform
is not limited in scope
to the image and video compression context.

\section*{Acknowledgments}

This work was partially supported by
CPNq, FACEPE, FAPERGS and FIT/UFSM (Brazil),
and by the College of Engineering at the University of Akron, Akron, OH, USA.

{\small
\bibliographystyle{IEEEtran}
\bibliography{dctref-clean}
}

\end{document}